\documentclass{nime-alternate}

\begin{document}
%
\conferenceinfo{NIME'15,}{May 31-June 3, 2015, Louisiana State Univ., Baton Rouge, LA.}

\title{Tracking Ensemble Performance on Touch-Screens with Gesture
  Classification and Transition Matrices} 

\numberofauthors{3} 
\author{
\alignauthor
Charles Martin\\
       \affaddr{Australian National University}\\
       \affaddr{Canberra, Australia}\\
       \email{charles.martin@anu.edu.au}
\alignauthor
Henry Gardner\\
       \affaddr{Australian National University}\\
       \affaddr{Canberra, Australia}\\
       \email{henry.gardner@anu.edu.au}
\alignauthor
Ben Swift\\
       \affaddr{Australian National University}\\
       \affaddr{Canberra, Australia}\\
       \email{ben.swift@anu.edu.au}
}
\maketitle

\begin{abstract}
  We present and evaluate a novel interface for tracking ensemble
  performances on touch-screens. The system uses a Random Forest
  classifier to extract touch-screen gestures and transition matrix
  statistics. It analyses the resulting gesture-state sequences across
  an ensemble of performers. A series of specially designed iPad apps
  respond to this real-time analysis of free-form gestural
  performances with calculated modifications to their musical
  interfaces. We describe our system and evaluate it through
  cross-validation and profiling as well as concert experience.
\end{abstract}


\acmclassification1998{H.5.5. [Information Interfaces and
  Presentation] Sound and Music Computing --- Systems, H.5.3. [Information Interfaces and
  Presentation] Group and Organization Interfaces --- Synchronous interaction}

\section{Introduction}

\begin{figure}[htbp]
\centering
\includegraphics[width=1\columnwidth]{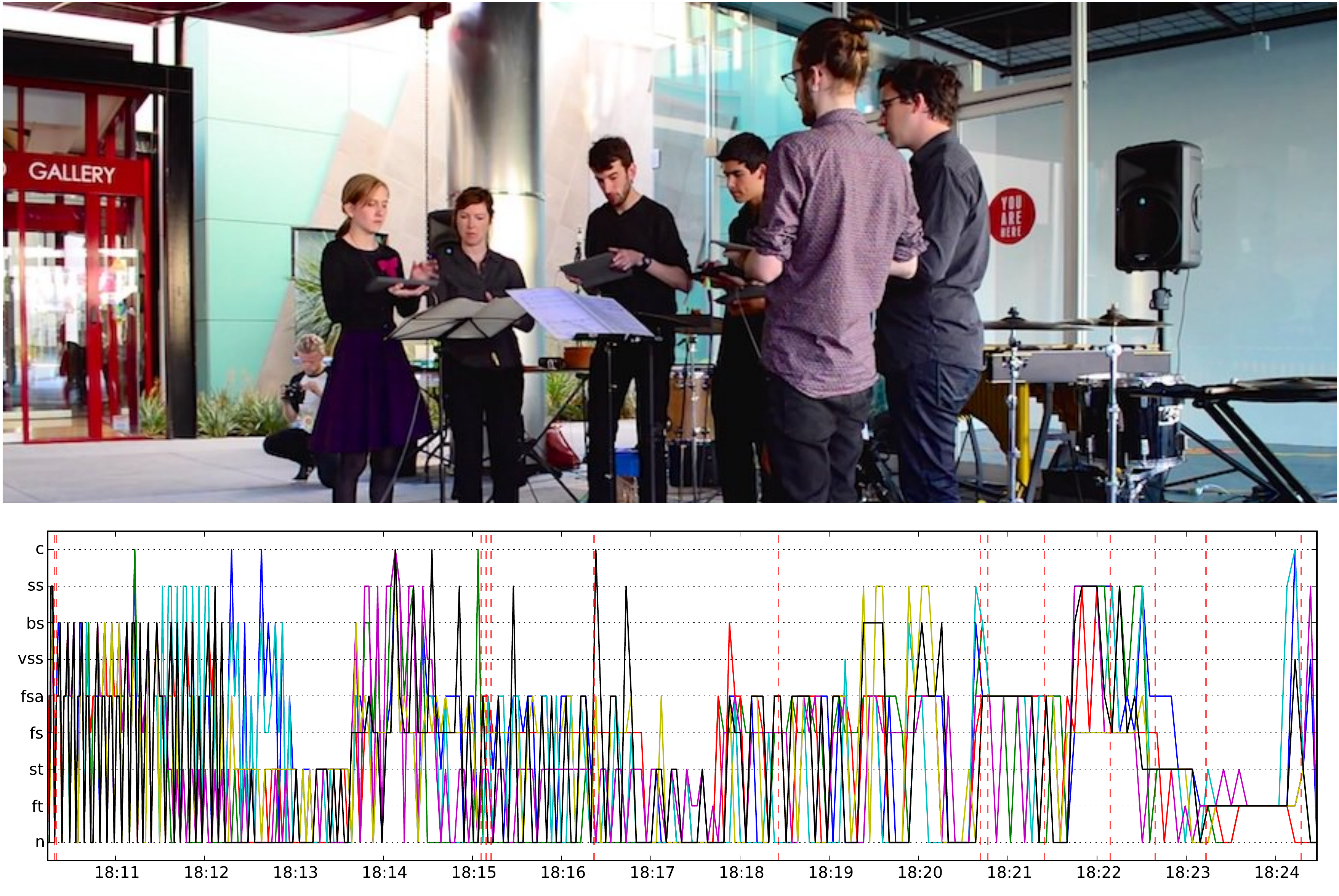}
\caption{An iPad septet performing with Metatone Classifier and our
  apps. The lower plot shows the ensemble's gestures over the whole
  performance. Moments of peak change that triggered ``new-idea''
  messages are marked with vertical red lines. }
\label{gesturestudy-graphoutput}
\end{figure}

Free-improvised ensemble musical performances can be considered as
sequences of musical sections segmented by moments where the group
spontaneously moves to explore a new musical
idea~\cite[pp.~58--59]{Stenstrom:2009xy}. This paper describes the
design and evaluation of a server-based agent that tracks this type of
musical interaction on touch screen apps and makes calculated
adjustments to the performers' interfaces based on the ensemble
performance to support their improvisational creativity.

Previous work has identified a vocabulary of gestures used by expert
percussionists on iPad interfaces~\cite{Martin:2014cr}. We used these
results to construct an agent that observes performers' touch-screen
interactions in real-time and classifies them as a sequence of
gestural states. Our agent estimates the occurrence of new ideas across
the ensemble by calculating a measure, flux, on the transition matrix
of these gesture states.

We have developed several iPad apps that are designed to respond to
this agent by updating their user interface. The aim with our apps is
to present an ``interface-free interface'' to the performers, where
the ensemble's musical direction is used to adjust pitches, effects,
and sonic-material available to the performers. Three of our apps will
be described in this paper that have different paradigms for
interaction with the agent; \emph{Snow Music} supports the performers
with complementary sounds when they continue certain gestures;
\emph{PhaseRings} rewards the performers with new pitches and
harmonies when they explore different gestures together; and
\emph{BirdsNest} disrupts performers who stay on certain gestures too
long with changes in the app's sound and features.

Evaluation of our agent's classifier has demonstrated a 97\% level of
accuracy when trained and evaluated with high-quality data. Time
profiling for a typical performance has shown that our system should
scale for use in live concerts with up to 25 performers. Experience
with our apps over several concerts demonstrates that the system is
practical and that our range of iPad apps provides performers with
opportunities to develop styles of gestural and musical interaction,
both with the agent and each other. In the next section, we will
discuss prior research in this area. Following that, we will describe
the construction of our system of agent and apps in detail and report
on the results of our evaluations.

\section{Background}

A common design pattern for computer music performances is the
``Laptop Orchestra''~\cite{TruemanLOrk,Bukvic:2010lq} (LO) where
multiple performers use similar hardware and software setups in an
ensemble performance. The formalism inherent in such groups allows the
establishment of compositional repertoire, a
pedagogy~\cite{Wang:2008qq}, and an emphasis on liveness. Frequently,
the software setup itself is considered to be the ``composed'' aspect
of the musical work and the performers improvise their own
parts~\cite{Smallwood:2008qv}.

In LOs and other ensembles employing new interfaces for musical
expression, artificial-intelligence agents have been used as
improvisation partners or as ensemble
members~\cite{MartinAengus:2011sh}. Such agents may be designed to
imitate a particular musician~\cite{Vempala:2007rc}, or to follow a
broader style~\cite{Pachet:2003wd} using statistical models including
those based on Markov processes~\cite{Ames:1989fj}. A more abstract
role for an agent in LOs is as a ``virtual
conductor''~\cite{TruemanLOrk} which communicates with the ensemble
providing cohesive direction of broad musical intentions. When
conducting or performing agents respond to other members of the
ensemble, they may be tracking features extracted from audio
streams~\cite{Hsu:2007qq} or the output of machine-learning algorithms
applied to sensors~\cite{fiebrink2009metainstrument}.

Simultaneously with the development of LOs, the concept of ``mobile
music'' has gained currency~\cite{Gaye:2006qy}. Here, powerful mobile
devices such as smartphones and tablets have been co-opted as musical
instruments~\cite{jenkins2012ipad} due to the affordances of their
multiple sensors, their convenient form-factors and their growing
cultural importance~\cite{Tanaka:2010sp}. Inevitably, mobile device
ensembles have emerged, using phones to perform gamelan-like
sounds~\cite{Greg-Schiemer:2007mz} or to develop a repertoire of
sensor-based music~\cite{2008-icmc-mopho}. The proliferation of
multitouch devices has emphasised exploration of touch user
interfaces~\cite{oh2010evolving} and of developing mobile apps as a
classroom activity~\cite{Essl:2010qr}. As with Tangible User
Interfaces~\cite{Xambo:2013pd} mobile devices present new
opportunities for participation and creativity in musical ensembles
and both smartphones~\cite{Swift:2013xy} and
tablets~\cite{Martin:2014cr} have been used in improvised-music
ensembles. In fact, their widespread adoption has led to their use in
music education; Williams~\cite{Williams:2014wt} has reported that
these meta\--instruments suggest exploratory and collaborative modes
of music making in the classroom.

The concept of capturing gesture in performance is central to the NIME
field~\cite{Jensenius:2014ul}. While there are many systems for
classifying or tracking gestures~\cite{Caramiaux:2013} during
performances, these are generally focussed on the performance of
individual musicians. In this research we present a system that
classifies the gestures of a mobile-music ensemble simultaneously and
continuously analyses the whole ensemble's behaviour. One approach for
analysing performer behaviour is to construct transition matrices of
changes between a set of musical states which characterise that
performance. This approach has been first described by Swift et~al in
their analysis of ``live coding'' protocols~\cite{Swift:2014tya}. In
the present work, we further develop this transition matrix approach
for real-time gestural analysis of touch-screen ensembles.

\section{System Design}

\begin{figure*}[htbp]
\centering
\includegraphics[width=1\textwidth]{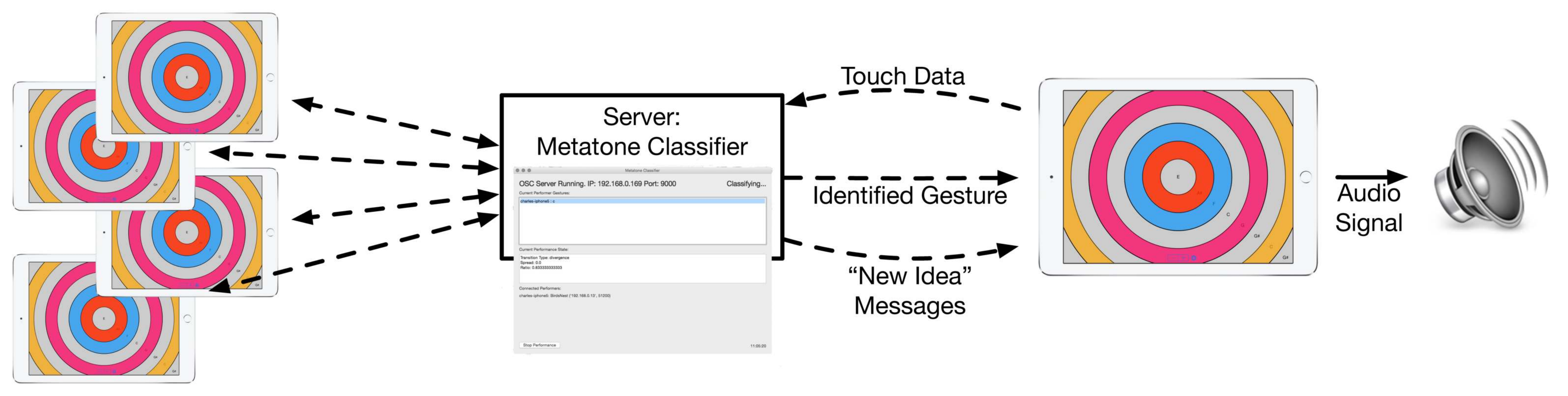}
\caption{The performance architecture of our system of server-based
  agent and iPad apps. Each iPad connects automatically to the server
  over WiFi. All touch interactions are logged and classified into
  gestures by the server which returns individual gesture and ensemble
  ``new-idea'' events throughout the performance. Each iPad's sound is
  projected from a loudspeaker via the iPad's headphone jack.}
\label{metatone-architecture}
\end{figure*}

Our agent, \emph{Metatone Classifier}, consists of a Python
application which can run on a laptop computer in the performance
venue, or on a remote server. The agent has no sound-output
capabilities and interacts with specially designed iPad apps which are
used by our ensemble as performance instruments. During performances,
the ensemble's iPad apps connect to the server over a Wi-Fi network
using Bonjour (zero-configuration networking) provided by the
\texttt{pybonjour} module. Once connected, the iPad apps send logs of
each touch event to the agent using the OSC message
format~\cite{osc-nime2009}. Once touch events have been sent to the
agent, it begins to analyse the performance and return information to
the performers' iPads at a rate of once per second. The analysis is
performed in two stages: firstly, each performer's recent touches are
classified into a gesture class which are returned to their iPad;
secondly, gesture transitions from the whole ensemble are compiled
into a matrix which can be analysed to measure the state of the whole
ensemble. This information is then sent to every iPad. While the agent
is generally operated as a server process, we have also developed a
simple UI for Apple OS X that allows the server to be monitored during
our research performances.

\subsection{Gesture Classifier}

\begin{table}
    \begin{tabular}{|l|l|l|l|}
    \hline
    \# & Code & Description & Group \\ \hline
    0 & N   & Nothing & 0 \\
    1 & FT  & Fast Tapping & 1\\
    2 & ST  & Slow Tapping & 1\\
    3 & FS  & Fast Swiping & 2\\
    4 & FSA & Accelerating Fast Swiping & 2\\
    5 & VSS & Very Slow Swirling & 3\\
    6 & BS  & Big Swirling & 3\\
    7 & SS  & Small Swirling & 3\\
    8 & C   & Combination of Swirls and Taps & 4\\ \hline
    \end{tabular}
    \caption{Touch-screen gestures that our classifier is trained to
      identify during performances.
      \label{tab:touchScreenGestures}}
\end{table}

\emph{Metatone Classifier} classifies gestures by calculating
descriptive statistics from each performer's touch data using a
sliding window of five~seconds duration. These include: frequency of
movement, frequency of touch starts, mean location of touches,
standard deviation of touch location and mean velocity. A Random
Forest classifier~\cite{Breiman:2001kx} from Python's
\texttt{scikit-learn}~\cite{scikit-learn} package was trained using
known examples of nine touch-screen gestures (see Table
\ref{tab:touchScreenGestures}) recorded in a studio session by our app
designer. At a rate of once per second during performances, the
classifier identifies each performer's gesture using the last five
seconds of collected touch-data. These timing parameters (five-second
windows reported once per second) were tuned by trial and error. The
server stores these identified gestures and also sends them to the
performers' iPads.

\subsection{Transition Matrices and Flux}
\label{sec:tm-and-flux}

Although classifications of each performer's current gesture is
useful, more interesting information about the performance can be
gained by analysing the performers' transitions between gestures.
Given a set of gestures $G$, each musician's gesture activity
can be represented as a sequence
\begin{equation}
  X_n \hskip 2em n = 1, \ldots, N
\end{equation}
where each $X_i$ is a member of $G$. To examine transitions between
gestures we can consider the sequence $X_n$ as a Markov chain, and
calculate its transition matrix. The transition matrix $P$ for a
Markov process with $m$ states is an $m \times m$ matrix
\begin{equation}
  p_{ij} = \Pr(X_{t+1} = j | X_t = i)
\end{equation}
such that the transition from state $i$ to state $j$ is given by the
entry in the $i^{th}$ row and $j^{th}$ column.

We can estimate the transition matrix for a whole performance. Let
$N_{ij}$ be the number of times that state $i$ was followed by state
$j$ in $X_n$. The maximum likelihood estimator of $P$ is then
\begin{equation}
  p_{ij} = \frac{N_{ij}}{\sum_j N_{ij}}
\end{equation}
The matrix $P$ is a concise way of characterising the gesture
transition behaviour of each musician's activity in the
performance~\cite{Swift:2014tya}. To summarise the whole ensemble's
activity, we can average the transition matrices of each performer.
The usefulness of this characterisation relies on the gesture
sequences being Markovian---that every transition explicitly depends
only on the previous state. While this assumption may be difficult to
justify over a long gesture sequences, our transition matrices are
calculated over short sections of 15 seconds length, which, arguably,
would encompass little long-term planning by the musicians.

To compare the ensemble activity between sections of the performance,
we derive a high level quantity, called flux, which measures how much
the musicians change gesture. The transition matrix can be interpreted
as a description of trajectories through the set of gestures, $G$. One
style of moving through this space is in a segmented fashion, where a
musician will spend long periods performing one gesture, only
occasionally changing to another. At the other end of the spectrum is
a more frantic approach where a musician jumps frequently between
gestures, never dwelling on any particular gesture for too long.

Mathematically, we can discriminate between these, and intermediate
styles of interaction by measuring the flux of the transition matrix
$P$, where
\begin{equation}
  \mathrm{flux}(P) = \frac{\|P\|_1-\|\mathrm{diag}(P)\|}{\|P\|_1}
\end{equation}
where $\|P\|_1 = \sum_{i,j}|p_{ij}|$ is the element-wise 1-norm of the
matrix $P$ and $\mathrm{diag}(P)$ is the vector of the main diagonal
entries of $P$.

The flux measure returns a value in the range [0,1]. It will return 0
when all non-zero elements of the matrix are on the main diagonal,
that is, the performers never change gesture, and return 1 when no
performer stays on the same gesture for two classifications in a row.
Flux is \emph{small} (closer to 0) when the ensemble rarely changes
gesture, and \emph{large} (closer to 1) when the performers change
gesture frequently and is, therefore, a measure of how quickly an
ensemble changes from state to state.

\subsection{Identifying ``New Ideas''}

In {\em Metatone Classifier}, we are particularly interested in
identifying moments of peak flux in the performance that might
correspond to performers exploring a new gestural idea. We want the
agent to report such moments to the performers' iPads so that they can
update their functionality in response. In our implementation, each
second, our agent computes the ensemble transition matrices of the two
previous 15 second windows of the performance, and calculates their
flux. When the most recent flux measurement exceeds the next most
recent by a certain threshold, the system reports a ``new-idea''
message back to the performers' iPads. As it is possible that a single
``new-idea'' would be captured by several sequential measurements, the
iPad apps include a rate-limiting function, that will ignore messages
arriving more frequently than once per minute. 



\subsection{A Repertoire of Touch-Screen Apps}

Three iPad apps, ``BirdsNest'', ``Snow Music'' and ``PhaseRings'',
have been designed to interact with our agent during performance. All
of these apps have sound material designed in Pure Data and integrated
into the app using \texttt{libpd}, with the remaining components
designed in Objective-C. The apps have a simple percussion-inspired
scheme for mapping touch to sound: they present users with a free-form
touch area for interacting with sample-based and pure synthesised
sounds. Tapping the screen produces a short, percussive sound while
swiping or swirling produces continuous sounds with a volume
proportional to the velocity of the moving touch point. Sound output
from the apps is via the iPads' headphone output which can be either
dispersed through a mixer and PA system or directly connected to
powered speakers. While the performers' interactions may be similar,
the apps have been designed to respond to gesture classifications and
new-idea messages according to three quite different paradigms:
BirdsNest has been designed to be {\em disruptive}, Snow Music to be
\emph{supportive} and PhaseRings to be \emph{rewarding}. In the
following sections we will describe each of our three apps in detail.

\subsubsection{BirdsNest}

BirdsNest allows performers to play with bird samples, field
recordings, and percussive sounds from a northern Swedish forest with
a backdrop of images from that location. The app consists of a
progession of four sonic scenes representing a journey from the forest
floor to a vantage point high in the trees. Each scene has a palette
of sound material of which only a few sounds are available to each
player. These sounds are triggered by tapping and swirling on the
backdrop image. The interface has a ``sounds'' button that performers
can use to shuffle sounds available to them from the available
palette. It also has a ``looping'' function controlled by a switch
where tapped notes are repeated approximately every five seconds for a
limited number of times with increasingly randomised pitch and rhythm.
Another switch controls an ``autoplay'' function where
field-recordings from the sound palette are generatively triggered as
a backing soundscape. Ensemble performance with BirdsNest consists of
an exploration through different palettes of sounds where progression
between the sonic scenes is entirely governed by the agent. BirdsNest
is designed to \emph{disrupt} the musicians' performance, to
discourage performers from staying on any one gesture for too long.
Based on gesture feedback from the agent, the app watches for runs of
identical gestures and responds by switching the looping and autoplay
features on or off in the user interface in order to prompt new
actions by the performers.

\begin{figure}[!t]
\centering
\includegraphics[width=0.8\columnwidth]{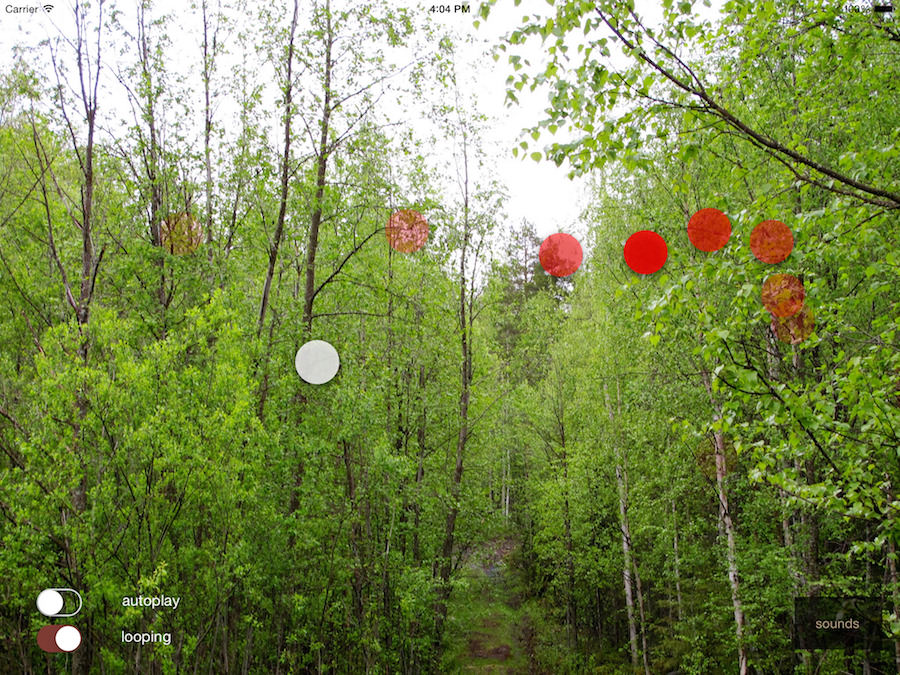}
\caption{The BirdsNest app is a sonic journey through field-recordings
  from a forest in Northern Sweden.}
\label{birdsnest-screenshot}
\end{figure}

\subsubsection{Snow Music}

\begin{figure}[htbp]
\centering
\includegraphics[width=0.8\columnwidth]{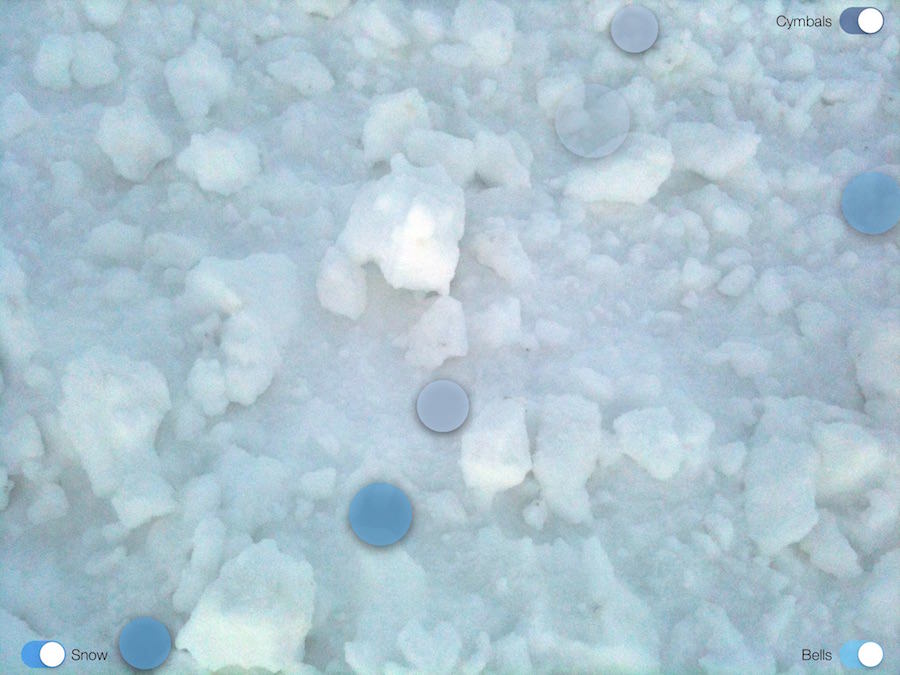}
\caption{Snow Music invites the player to manipulate samples of snow
  with their touch gestures. The UI buttons display the status of
  generative processes that support particular gestures. The blue
  circles represent notes triggered by these processes.}
\label{snowmusic-screenshot}
\end{figure}

Snow Music aims to emulate a bowl of snow, allowing performers to
manipulate recordings of snow being squished, smeared, stomped and
smashed. The app is designed so that performers can only unlock new
sounds or textures by interacting with these snow sounds, not by
activating UI elements. Snow Music uses a \emph{supportive} paradigm
for interaction with {\em Metatone Classifier}. The app watches for
runs of similar gestures and activates extra sounds that support the
player's intent. For instance, a run of tapping gestures causes the
app to layer the snow sounds with a glockenspiel sound when the user
taps while continuous swirling activates a generative backdrop of
melodic bell sounds. These supportive sounds are switched off when the
performer explores other gestures. In the case of Snow Music, new-idea
messages shuffle the snow samples available to the player and change
the pitches used in the supportive sounds. While the presence of the
supportive sounds are shown on the screen with UI switches and
animations, the performers are not able to control them directly with
UI elements. Although this app appears to have a limited selection of
sounds available to the player, the interaction with the agent
challenges the individual and the whole ensemble to fully explore a
range of touch-gestures together. The aim is to support mindful
exploration with a range of complementary musical elements.

\subsubsection{PhaseRings}

\begin{figure}[htbp]
\centering
\includegraphics[width=0.8\columnwidth]{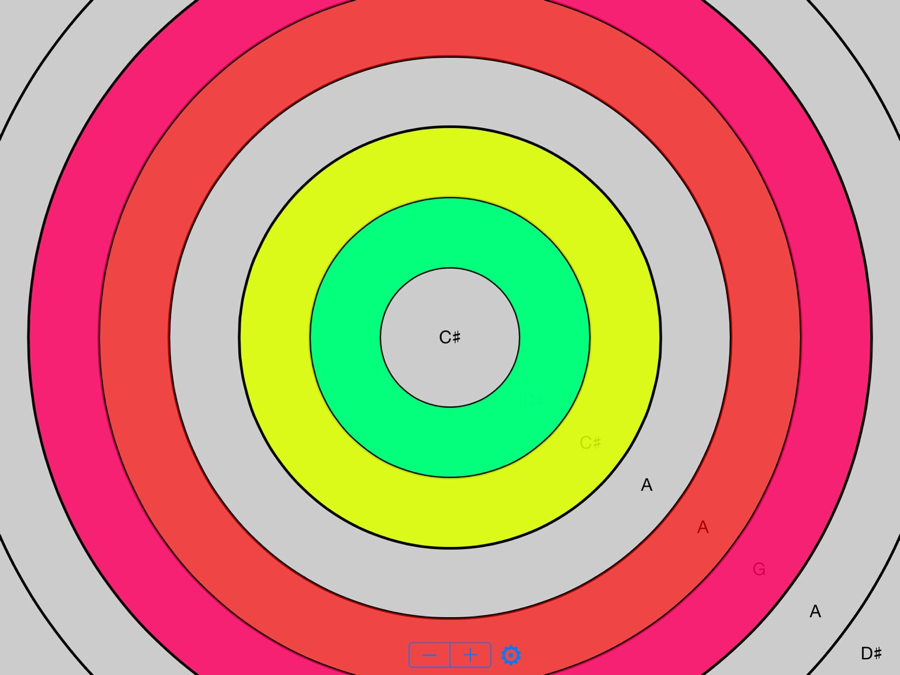}
\caption{The PhaseRings app where available
  notes are represented by rings that can be played by tapping or
  swirling.}
\label{singingbowls-screenshot}
\end{figure}

PhaseRings presents users with an abstract circular interface for
performing with percussive samples and pure synthesis sounds.
Concentric rings on the screen indicate where touches will activate
different pitches of a single sound source. Tapping a ring will
activate a note with a natural decay while swirling on a ring will
create a sustained sound. The PhaseRings app is configurable, through a
settings menu, to use one of seven different sound generators including
several percussive samples which sustain through granular-synthesis, a
phase-distortion sound, and a Karplus-Strong modelled string sound
sustained by tremolo. The app generates pitches randomly for each
player from a scale. Three scales are incorporated in the app as a
harmonic sequence with two setups of pitches from each scale (although
in ensemble performance the players cannot proceed through these
setups manually). PhaseRings \emph{rewards} the player's exploration of
gestures with these new pitches and harmonic material. When a new-idea
message is sent to the ensemble, the app changes the number and
pitch of rings displayed on screen. As the performers explore
different touch gestures, they are rewarded with the opportunity to
perform new melodic material with access to new notes and have a sense
of cohesive harmonic progression as the scale for each player's iPad
is uniform across the ensemble.

\section{System Evaluation}

\subsection{Gesture Classifier Accuracy}

{\em Metatone Classifier} was evaluated using standard
cross-val\-idation methods from machine learning. Three sets of training
data were available for a comparative evaluation: a proof-of-concept
set of example gestures with feature-vectors calculated on 5 second
windows, the same set using a rolling 5-second window at one-second
intervals, and a ``production set'' captured following a formal
procedure. The proof-of-concept gesture data was collected by matching
video analysis of a performance of examples of each gesture on one of
our iPad apps together with the logged touch data. The production set
was collected using a survey application, written in the Processing
environment\footnote{\url{http://www.processing.org}}, that instructed
the performer to play each gesture on an iPad app in randomised order
for one minute with a 20 second break in between each gesture. The
data was subsequently trimmed of the waiting periods and of ambiguous
frames at the beginning and end of each example gesture.

The three classifiers were evaluated using stratified 10-fold cross
validation which was performed 10 times on each training set,
producing 100 estimates of accuracy for each classifier. A one-way
ANOVA procedure revealed a significant effect of training set on
accuracy with $F(2,297) = 31.7, p < 0.001$. Paired
Bonferroni-corrected t-tests confirmed significant ($p < 0.05$)
differences between the three sets of training data with the newest
set producing a mean accuracy of 0.973 with standard deviation of
0.022. This level of accuracy is consistent with that reported in
other systems that recognise touch command
gestures~\cite{Wobbrock:2007kq}. It is notable that the production
training set produced a significantly more accurate classifier even
though the number of example gestures was only 9.5\% higher. The
improvement was more likely due to the quality of data collected using
our survey application.

\begin{table}
  \begin{tabular}{|l|l|l|l|l|}
    \hline
    \#& Description             & N   & Mean  & S.D. \\ \hline
    1 & 2013 gestures 5s window & 98  & 0.915 & 0.08 \\ 
    2 & 2013 gestures 1s window & 486 & 0.942 & 0.032\\ 
    3 & 2014 gestures 1s window & 532 & 0.973 & 0.022\\ \hline
  \end{tabular}
  \caption{We compared classifiers generated from three sets of
    feature vectors and known gestures. This table shows the number of vectors, mean
    accuracy and standard deviation.}
  \label{tab:trainingGestures}
\end{table}

\subsection{Computational Cost}

The computational cost of our agent was profiled using the
\texttt{line\_profiler}\footnote{\url{http://github.com/rkern/line_profiler}}
Python module during performances with zero to four iPad performers.
The test system ran Apple OS X on an Intel Core i7-2720QM 2.2GHz
processor. With four iPads, the most common configuration in our
performances, the classification and analysis function which is
triggered once per second took a mean time of 0.158s to complete. The
major components of this function were the calculation of feature
vectors for the performers (0.06s), the Random Forest classifications
(0.032s), and the calculation of transition matrices (0.049s).

The mean time for the classification and analysis function to complete
had a significant ($p < 0.001$) linear relationship with the number of
iPads performing. We can estimate from the linear model that on our
test system this function could take 0.038s per iPad plus 0.0085s
overhead. This suggests that an ensemble of around 25 iPads could be
an upper-bound for analysis in the desired one-second timeframe (with
similar hardware). Although this would be sufficient for the
membership of most institutional computer music ensembles, the
ubiquity of mobile-devices such as the iPad suggest that large scale
performances of much larger ensembles could be possible.


\subsection{Performance}

\emph{Metatone Classifier} was premiered in concert performance with
the three apps described in this paper in March 2014. Since then the
system has been used in several live performances with ensembles of
between two and seven performers as well as in an installation
context. 

In discussions conducted during rehearsals of an iPad quartet
(including one author of this paper), the performers reported
different reactions to the different modes of interaction with the
agent available in the three apps. They described their personal
styles for drawing out particular sounds through gestural interactions
with the agent, and their attempts to replicate their favourite
moments from rehearsals. This feedback has confirmed that our system
of apps and agent is practical for real-world performances and affords
creative and satisfying music-making.

\section{Conclusions}

We have presented a novel system for ensemble touch-screen musical
performance including a server-based agent that classifies performers'
gestures and tracks new ideas, and three iPad apps that use this
online agent to support, disrupt, and reward gestural exploration in
collaborative improvised performances. Our implementation of this
system uses a novel flux measure to determine change points in the
group's activity and to make calculated, real-time interventions to
the iPad interfaces. We have presented the results of an evaluation of
the gesture classifier used in the system where three training sets of
data were compared using cross-validation. We also profiled the
gesture classification and ensemble tracking algorithms in our system
to estimate an upper bound for ensemble size given our desired
time-frame of one analysis per second. The use of our system in a
series of concerts confirms that it is practical and supportive of
exploratory mobile-music performances.

While we have typically performed with four iPad players using this
system, the profiling results suggest that our current agent could
scale up to around 25 players before classifications would be delayed.
However, it is possible that timely response from the agent is not
critical and that a longer analysis cycle could be used to cater for
very large iPad ensembles.

The evaluation of our classifier revealed that training data collected
under controlled conditions had produced a significantly more accurate
Random Forest Classifier even though the size of the training sets
were similar. This result justifies the extra effort required to
design a system for automatically and accurately capturing training
data rather than the previous manual analysis of video-recorded
gestures.

There are several ways that the research described here can be
extended in the future. While the computational cost of performing a
gesture classification and generating transition matrices will
increase with the size of an ensemble, the cost of measures on the
transition matrix will not since it has a fixed size. Other matrix
measures may reveal different aspects of the ensemble's musical
behaviour and could probably be incorporated without a significant
cost in computation. While the agent has been used with co-located
performers, our future goal is to use the system in
networked performances where the agent responses may assist
the performers' feelings of cohesion with remote participants.






\bibliographystyle{abbrv}
\bibliography{nime2015-references}  
\end{document}